\documentclass[a4paper]{jpconf}
\usepackage{graphicx}
\begin{document}
\title{The Jefferson Lab 12 GeV Upgrade}

\author{R.~D.~McKeown}

\address{Thomas Jefferson National Accelerator Facility, Newport News, Virginia 23606, USA}
\address{Department of Physics, College of William and Mary,
Williamsburg, Virginia 23187, USA}

\ead{bmck@jlab.org}

\begin{abstract}
Construction of the 12~GeV upgrade to the Continuous Electron Beam
Accelerator Facility (CEBAF) at the Thomas Jefferson National
Accelerator Facility is presently underway. This upgrade includes
doubling the energy of the electron beam to 12~GeV, the addition of
a new fourth experimental hall, and the construction of upgraded
detector hardware. An overview of this upgrade project is presented,
along with highlights of the anticipated experimental program.
\end{abstract}

\section{Introduction}

Since 1995, the CEBAF facility at Jefferson Laboratory has operated
high-duty factor (continuous) beams of electrons incident on three
experimental halls, each with a unique set of experimental
equipment. As a result of advances in the performance of novel
superconducting radiofrequency (SRF) accelerator technology, the
electron beam has exceeded the original 4~GeV energy specification,
and beams with energies up to 6~GeV with currents up to 100~$\mu$A
have been delivered for the experimental program. In addition, the
development of advanced GaAs photoemission sources has enabled high
quality polarized beam with polarizations up to 85\%. The facility
serves an international scientific user community of over 1200
scientists, and to date over 160 experiments have been completed.

Shortly after the first experiments were started, planning began to
upgrade the capability of this facility to enable beams up to 12~GeV
in energy. In 2002, the Long Range Plan of the US Nuclear Science
Advisory Committee (NSAC) \cite{lrp02} contained a
recommendation:``We strongly recommend the construction of CEBAF at
Jefferson Laboratory to 12 GeV as quickly as possible''. In March
2004 the US Department of Energy (DOE) granted CD-0 approval to
develop a conceptual design for such a facility. In 2007, the NSAC
Long Range Plan \cite{lrp07} reaffirmed the community commitment to
this project: ``We recommend completion of the 12 GeV Upgrade at
Jefferson Lab.  The Upgrade will enable new insights into the
structure of the nucleon, the transition between the hadronic and
quark/gluon descriptions of nuclei, and the nature of confinement.''
Considerable effort by the Jefferson Lab user community and the
Laboratory staff facilitated DOE final approval for the start of
construction in September 2008.

\section{Description of Construction Project}

Currently, CEBAF is a recirculating linac, with 2 linac sections
consisting of 20 cryomodules each, each cryomodule containing 8
superconducting RF cavities. These cryomodules are each capable of
an average 25~MeV of acceleration, nominally producing 0.5~GeV
acceleration in each linac section. The recirculating arcs contain
quadrupole and dipole magnets in separate beamlines that enable the
beam to be accelerated up to 5 times through both linacs, producing
a nominal energy of 5~GeV with actual performance up to 6~GeV. After
the second (south) linac, beam pulses can be "kicked" into an
extraction line into one of the 3 experimental halls. Thus the
1497~MHz microstructure can be split into three 499~MHz beams with
energies in multiples of $\frac{1}{5}$ of the full 5-pass energy.

\begin{figure}
\begin{center}
\includegraphics[width=\columnwidth]{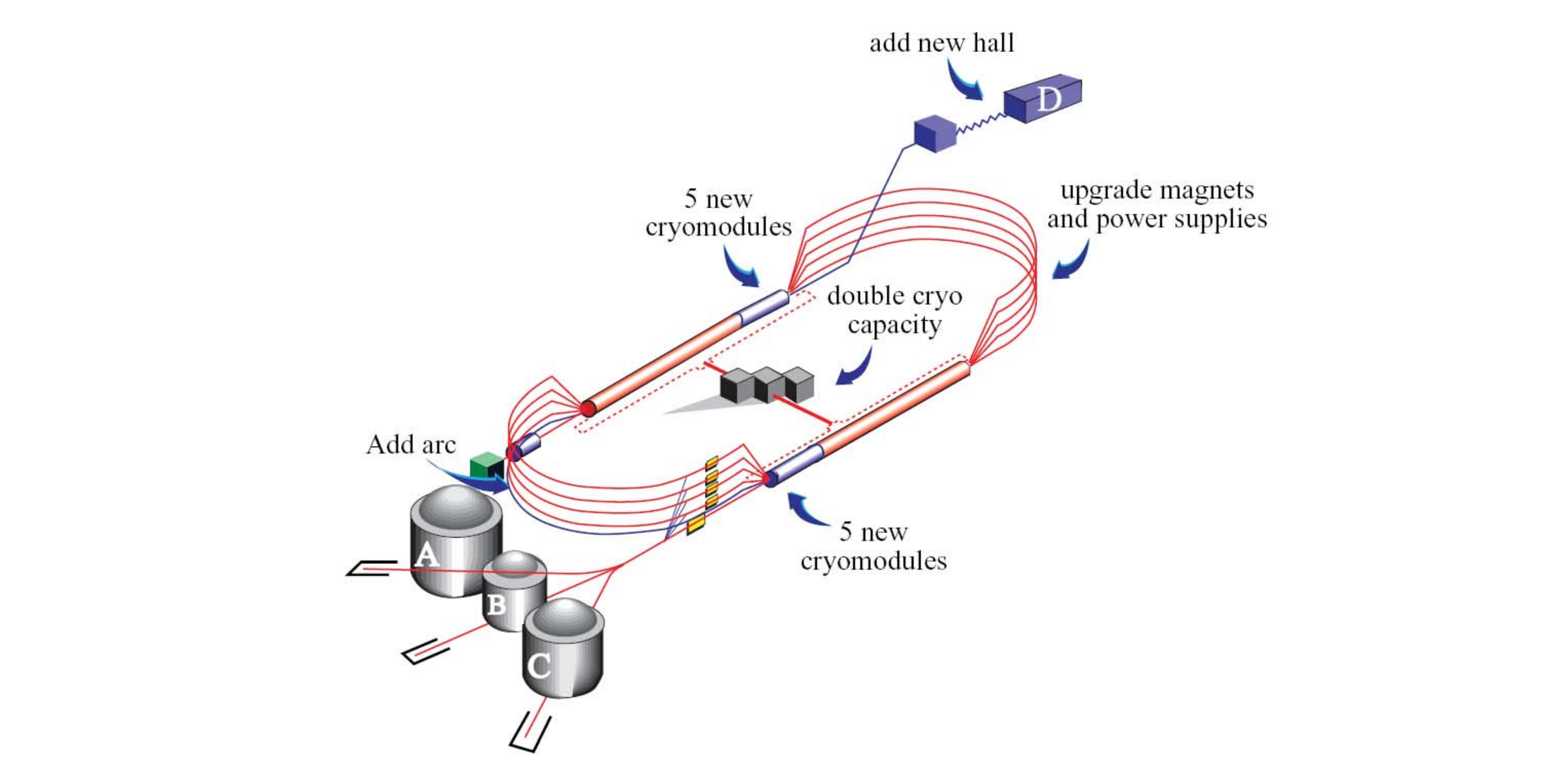}
\end{center}
\caption{\label{fig:overview}Jefferson Lab 12 GeV upgrade concept.}
\end{figure}

The basic concept of the 12 GeV upgrade project is illustrated in
Fig.~\ref{fig:overview}. In addition to the upgrade of the
accelerator system to enable delivery of 12~GeV beam, the
experimental equipment will be enhanced to facilitate full
exploitation of the higher energy beam. This includes substantial
new equipment in Hall B and Hall C, and a completely new Hall D with
a new detector spectrometer system. The plan for Hall A also
includes upgraded and new equipment that is outside the present
construction project.

\subsection{Accelerator system upgrade}

The heart of the accelerator upgrade design is the addition of 5
additional cryomodules in each linac section. The original CEBAF
construction project left extra drift space at the end of each linac
section, enabling an energy upgrade. The new cryomodules will
contain 7-cell cavities instead of the 5-cell cavities in the
original CEBAF. In addition, improved surface treatments now enable
higher RF fields. The result is that each cryomodule will be capable
of 100~MeV of acceleration (compared to 20~MeV in the original
design specification). Thus each linac will now be capable of
1.1~GeV, and each recirculated beam will be able to reach 2.2~GeV.
With upgrades to the arc magnets and power supplies, we would be
able to deliver beams to all 3 existing experimental halls at
energies up to 11~GeV, with beam available in multiples of
$\frac{1}{5}$ of the full 5-pass energy. For example at full energy,
beam would be available at three of the values corresponding to
2.2~GeV, 4.4~GeV, 6.6~GeV, 8.8~GeV, and 11~GeV. The intensity of
these beams may total up to $85~\mu$A with high polarization (up to
85\%) available.

A fifth arc of magnets will be added to the 2nd recirculating arc,
so that one beam can be accelerated through the north linac one more
time (to reach a total of 12~GeV before transmission to the new
experimental Hall D. In addition, the central helium liquifier
capacity is being doubled to provide increased delivery of liquid
helium.

\subsection{Hall A upgrade}

Hall A will retain the present capability with 2 High Resolution
Spectrometers (HRS), and will host new experimental equipment to be
funded outside the present scope of the 12~GeV upgrade project. As
an example, the proposed Super Big-Bite Spectrometer (SBS) will be
sited in Hall A. This is a larger version of the recently deployed
Big-Bite Spectrometer. The SBS will consist of a large 48D48 dipole
magnet followed by large area tracking (GEM detectors),
time-of-flight scintillators,  and particle ID detectors (Cerenkov
and calorimeter). The solid angle acceptance depends on the
deployment geometry but is typically 50-100~msr, and the
spectrometer will be capable of operation at high luminosities
exceeding $10^{38}$/cm$^2$-s.

Hall A will also be the site of future high-precision parity
violation experiments to provide stringent tests of the standard
electroweak model. Measurements of parity-violating deep inelastic
scattering will be pursued with the proposed solenoidal detector
SOLID in Hall A.

\begin{figure}
%\begin{minipage}{14pc}
\begin{center}
\includegraphics[width=\columnwidth]{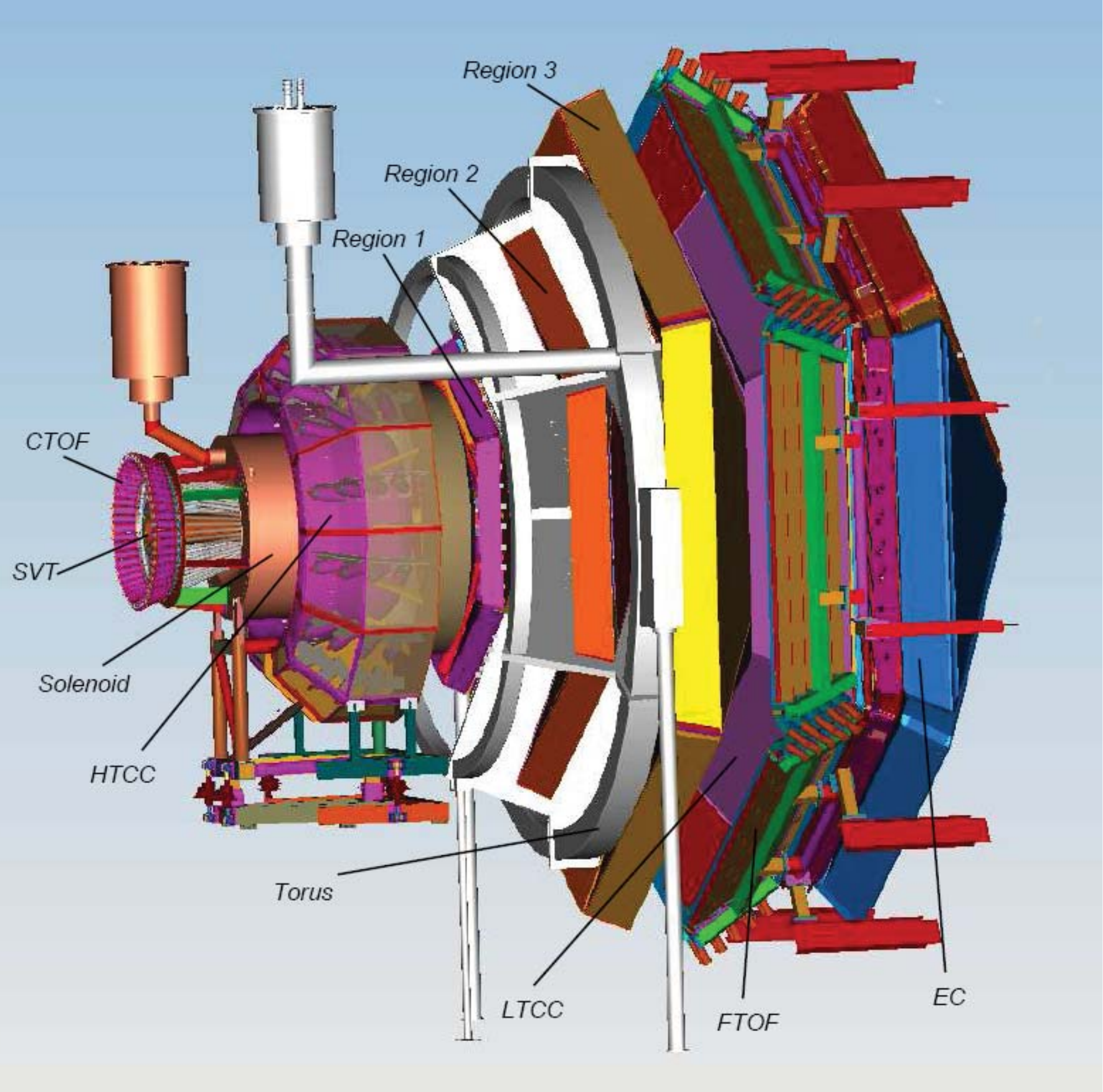}
\caption{\label{fig:CLAS12}3D view of CLAS12. The beam is incident from the left. The target is located inside the solenoid magnet.}
\end{center}
\end{figure}

\subsection{Hall B upgrade}

Hall B was the home of the CEBAF Large Acceptance Spectrometer
(CLAS). This apparatus will be replaced by a new spectrometer,
CLAS12, shown in Fig.~\ref{fig:CLAS12}. CLAS12 is designed to enable
higher luminosity operation with electron beams, up to
$10^{35}$/cm$^2$-s, and improved particle identification at forward
angles. The magnetic system includes a 5~T superconducting solenoid
for central tracking, and a superconducting toroid at forward
angles. The high field solenoid also shields the detector system
from the high flux of M{\o}ller electrons generated in the target.

The forward detector system includes a high threshold cerenkov
counter (HTCC) with pion threshold of 4.9~GeV mounted in front of
the toroid. The toroid and forward drift chamber system are designed
to track charged particles in the range $5-40^\circ$, and are
followed by a low threshold cerenkov counter (LTCC). The existing
electromagnetic calorimeter (EC) is being upgraded with a pre-shower
counter. The central detector system includes a silicon vertex
tracker and a scintillator time-of-flight array.

\subsection{Hall C upgrade}

In Hall C, the existing high momentum spectrometer (HMS) will be
complemented with a new super high momentum spectrometer (SHMS). The
SHMS design is a QQQD with a horizontal bend, capable of a maximum
central momentum of 11~GeV with -11 to +22\% momentum acceptance.
The solid angle acceptance is 5~msr and the momentum resolution will
be better than $10^{-3}$. The SHMS detector package includes a drift
chamber system for tracking, quartz and scintillator hodoscopes, a
noble gas Cerenkov counter and a heavy gas Cerenkov counter, and a
lead glass calorimeter.

\begin{figure}
\begin{center}
\includegraphics[width=\columnwidth]{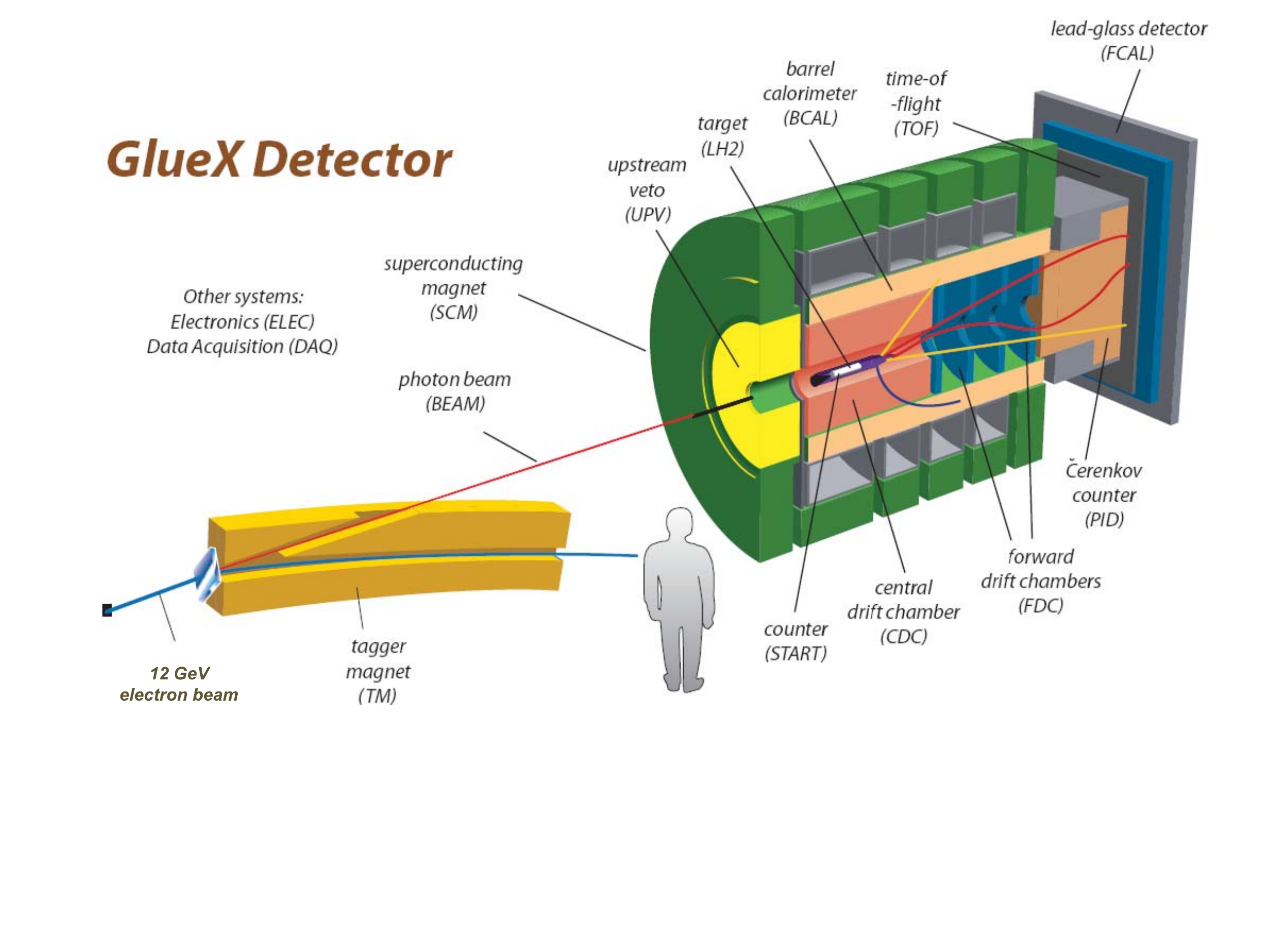}
\end{center}
\vspace{-1.0in}
\caption{\label{fig:gluex}Schematic of GlueX apparatus.}
\end{figure}

\subsection{Hall D}

The new experimental Hall D will be the host of the GlueX experiment
(see Fig.~\ref{fig:gluex}). As discussed below, the main goal of the
Gluex experiment is to search for exotic mesons produced in
photoproduction on the nucleon. Hall D will be preceded by a photon
tagging system. Linearly polarized photons will be produced by
coherent bremmstrahlung on a thin ($\sim 20$ micron) diamond wafer.
The scattered electrons from 8.5-9~GeV photons will be tagged with
scintillator detectors following a bending magnet, yielding a tagged
photon resolution of 0.2\% with expected fluxes to reach $10^8$/s.

The Gluex detector system includes a superconducting solenoid with
2.2~T field over a length of 495~cm and with an inner bore of
185~cm. The 30 cm-long liq- uid hydrogen target is surrounded by
scintillation counters (START), a cylindrical drift chamber array
(CDC) and an electromagnetic lead/scintillating fiber calorimeter
with a barrel geometry (BCAL). Downstream of the target are an array
of planar drift chambers (FDC). Outside and downstream of the clear
bore of the magnet are a Cerenkov counter, a wall of scintillation
counters (TOF) to measure time-of-flight and an electromagnetic
calorimeter (FCAL) consisting of lead-glass blocks. The angular
coverage will be from $1-120^\circ$ for charged particles, and from
$2-120^\circ$ for photons.

\section{Physics Program}

The physics program to be addressed with the Jefferson Lab 12~GeV upgrade has been developed in collaboration with the user
community and with the guidance of the Program Advisory Committee. There are presently 34 approved experiments, and 13
additional proposals have conditional approval. These proposals fall into four general categories:
\begin{itemize}
  \item the physical origins of quark confinement (including meson and baryon spectroscopy),
  \item the spin and flavor structure of the proton and neutron,
  \item the quark structure of nuclei,
  \item discovery of new physics through high precision tests of the Standard Model.
\end{itemize}
It is not possible here to provide a comprehensive review of this broad program, so a few selected highlights will be presented.

\subsection{Meson Spectroscopy with GlueX}

The GlueX experiment \cite{gluex} will provide unprecedented
capability to study the spectrum of light quark mesons in
photoproduction. A major goal is to search for mesons with exotic
quantum numbers ($J^{PC}= 0^{+-}, 1^{-+}, 2^{+-}$) that cannot be
described by quark-antiquark states. One can construct such states
in models \cite{isgur} that incorporate degrees of freedom
associated with excitation of a gluonic flux tube connecting the
quark and antiquark. More recently, these states and their
properties have been studied in detail using lattice QCD methods
\cite{LQCD}, and the lattice calculations support the interpretation
of excitation of gluonic degrees of freedom.

\subsection{Nucleon Spin Structure}

The study of the spin structure of the nucleon is still a very
active area of research. For the last 25 years, studies of the spinm
-dependent parton distributions have indicated that the quark
helicities contribute a small fraction of the nucleon spin
\cite{Bass}. Recent data from RHIC and semi-inclusive Deep Inelastic
Scattering (SIDIS) indicate that the gluon spin contribution is also
small \cite{DSSV}.  So although there has been considerable progress
in constraining the spin-dependent parton distribution functions, a
satisfactory accounting of the angular momentum of the nucleon in
terms of its fundamental constituents is still lacking. It now
appears that the orbital angular momentum of the quarks may play a
substantial role, and new experimental efforts are being proposed to
address this issue. There has been a great deal of theoretical
progress in the development of experimental observables that could
reveal aspects of the quark orbital angular momentum. These include
generalized parton distributions (GPD) that can be accessed through
deeply virtual Compton scattering (DVCS) \cite{GPD_Ji, GPD_Rad}, and
transverse momentum distributions \cite{TMD} which can be studied in
semi-inclusive deep inelastic scattering (SIDIS). The large
acceptance available with the CLAS12 detector system makes it
well-suited to the study of DVCS \cite{CLAS12}. Measurements of TMD
with SIDIS will be an important topic for the proposed SOLID
spectrometer in Hall A.

\subsection{Testing the Standard Model}

The discoveries of dark matter, dark energy, and the flavor
oscillations of neutrinos (associated with their small but finite
masses) are all indications that the standard model of the strong
and electroweak interactions requires modification. In addition,
there are theoretical motivations for extending the standard model
associated with protecting the Higgs mass from uncontrolled loop
corrections. These issues generally lead to the view that the
standard model is part of a larger theoretical framework, and such
an extension of the theory should lead to observational
consequences. Thus there are experimental activities associated with
two general methods for addressing this issue. One is to advance the
high energy frontier with higher energy particle accelerators to
attempt to obtain evidence for new particles that can be produced at
higher energy. The other method is to perform high precision
measurements at lower energies to provide information that can
indicate the properties of a more complete theory. Indeed, such
precision measurements yield complementary information to that which
can be obtained at the energy frontier. For a review of this
approach, including discussion of parity-violating electron
scattering, see \cite{MJRM}.

The 12~GeV upgrade of CEBAF will offer new opportunities for
precision measurements, particularly in studies of parity violating
electron scattering. The neutral weak interaction violates parity
symmetry, and so measurement of parity violation in electron
scattering is a consequence of interference of neutral weak and
electromagnetic amplitudes. Extensions of the standard model may
involve the existence of new massive ($M> M_Z$) neutral bosons that
couple to electrons and/or quarks. Such phenomena, or other similar
processes, would lead to changes in the parity-violating interaction
with electrons and thus manifest themselves in parity violation
experiments.

There are presently 2 proposals to perform such measurements at the
upgraded CEBAF. One would use the SOLID solenoidal magnetic
spectrometer system to study parity-violating deep inelastic
scattering \cite{SOLID}. The other proposal involves the
construction of a novel dedicated toroidal spectrometer to study
parity-violating M{\o}ller scattering \cite{Moller}. Both
experiments propose to be sited in experimental Hall A.

\section{Summary}

The 12~GeV upgrade project at Jefferson Lab will enable a powerful
new experimental program that will advance our understanding of the
quark/gluon structure of hadronic matter, the nature of Quantum
Chromodynamics, and the properties of a new extended standard model
of particle interactions. Commissioning of the upgraded beam will be
begin in 2013, and the full complement of upgraded experimental
equipment will be completed in 2015. This unique facility will
provide many opportunities for exploration and discovery for a large
international community of nuclear scientists.

\section{Acknowledgements}
I am grateful to Allison Lung, Kees de Jager, Volker Burkert, Steve
Wood, and Eugene Chudakov for their assistance in preparing this
presentation. This work was supported by DOE contract
DE-AC05-06OR23177, under which Jefferson Science Associates, LLC,
operates the Thomas Jefferson National Accelerator Facility.

\end{document}